\documentclass[12pt]{article}

\usepackage[left=1in,top=1in,right=1in,bottom=1in,nohead,paperwidth=8.5in, paperheight=11in]{geometry} 
\usepackage{amsmath,amsfonts,amssymb,amsthm,dsfont }
\usepackage{geometry}
\usepackage{xcolor}
\usepackage{mathrsfs}
\usepackage{graphicx}
\usepackage{enumerate}
\usepackage{algpseudocode}
\usepackage{bbm}
\usepackage{hyperref}
\usepackage{float}
\usepackage{caption}
\usepackage[toc]{appendix}
\usepackage{algorithm}
\usepackage{setspace}
\usepackage{booktabs}
\usepackage{longtable}
\usepackage{natbib}

\title{\bf Regularized Predictive Models for Beef Eating Quality of Individual Meals}
\author{Garth Tarr$^{a}$\footnote{Corresponding author. E-mail: garth.tarr@sydney.edu.au, i.wilms@maastrichtuniversity.nl. \newline
Acknowledgments: The authors are very grateful to Meat and Livestock Australia for providing the access to the data set. Garth Tarr was supported by Meat and Livestock Australia (L.EQT.2201), the Australian Research Council (DP210100521) and the AIR$@$innoHK program of the Innovation and Technology Commission of Hong Kong. 
} and Ines Wilms$^{b}$
		\\ \textit{\small $^{a}$ School of Mathematics and Statistics, The University of Sydney, Australia }
	\\ \textit{\small $^{b}$ Department of Quantitative Economics, Maastricht University, Maastricht, The Netherlands}
}
\date{ }

\begin{document}

\def\spacingset#1{\renewcommand{\baselinestretch}%
	{#1}\small\normalsize} \spacingset{1}
	
\maketitle

\vspace{-12pt}
\noindent
{\bf  Abstract.}
Faced by changing markets and evolving consumer demands, beef industries are investing in grading systems to maximise value extraction throughout their entire supply chain.  
The Meat Standards Australia (MSA) system is a customer-oriented total quality management system that stands out internationally by predicting quality grades of specific \textit{muscles} processed by a designated \textit{cooking method}. 
The model currently underpinning the MSA system requires laborious effort to estimate and its prediction performance may be less accurate in the presence of unbalanced data sets where many ``muscle$\times$cook'' combinations have few observations and/or few predictors of palatability are available.   
This paper proposes a novel predictive method for beef eating quality that bridges a spectrum of muscle$\times$cook-specific models. 
At one extreme, each muscle$\times$cook combination is modelled independently; at the other extreme a pooled predictive model is obtained across all muscle$\times$cook combinations. 
Via a data-driven regularization method, we cover all    muscle$\times$cook-specific models along this spectrum. 
We demonstrate that the proposed predictive method attains considerable accuracy improvements relative to independent or pooled approaches on unique MSA data sets. \\
\bigskip


\newpage
\spacingset{1.5} 

\section{Introduction} \label{sec: Introduction}
Beef industries around the world have been investing heavily in grading systems to create consistency and transparency in red meat supply chains.
Such grading systems are important drivers of efficiency improvement and customer awareness across the entire beef production chain. 
Upstream, knowledge on important palatability predictors is crucial to improve breeding and processing outcomes by, for instance, delivering cattle at more desirable weight, age and fatness endpoints to produce high-quality meat (e.g., \citealp{berri2019predicting, polkinghorne2008current}).
Downstream, accurate eating quality predictions are vital to devise differential pricing strategies since consumers are willing to pay premiums for higher quality products, or to install marketing campaigns allowing retailers to market muscles under cooking styles that optimize their quality (e.g., \citealp{polkinghorne2008}).

We focus on the Meat Standard Australia (MSA) beef grading system \citep{watson2008development} whose effectiveness has been proven internationally  (e.g., \citealp{bonny2018update}).
Many international beef grading systems, such as USDA beef grading system in America or the EUROP grid system in Europe, assign a single score to each carcase (e.g., \citealp{elliesoury2020various}). 
In contrast, the MSA beef grading system is a statistical model designed to predict the degree of customer satisfaction for an individual meal \citep{polkinghorne2008current}. 
Rather than describing an entire carcass, MSA beef eating quality predictions are generated for individual \textit{muscles} for each \textit{cooking method}. 
We refer to this as modelling \textit{``muscle$\times$cook"} combinations.
To predict eating quality of beef for each muscle$\times$cook combination, testing was widened to evaluate multiple cooking methods on an expanded range of muscles. 
Alternative cooking methods often generate widely different scores for the same muscle; while various indicators of palatability impact individual muscles differently \citep{polkinghorne2018commodity}. 
Predictive models that pool across muscles and/or cooking methods or that produce a single carcass score are thus inappropriate. 
Instead, a muscle$\times$cook-specific predictive modelling approach is needed.

Such a muscle$\times$cook-specific prediction perspective generates many statistical challenges.  
These challenges mainly arise due to the unbalanced nature of the data set.
Some muscle$\times$cook combinations, such as grilled striploins, have many recorded observations as it is considered a benchmark muscle and cooking method. 
As the majority of experiments conducted include grilled striploins we are able to obtain sufficient variation in palatability predictors to ensure our model estimates are a valid representation of the underlying data generating mechanism and obtain accurate predictions for this muscle$\times$cook combination. 
However, many muscle$\times$cook combinations have far fewer recorded observations and/or the observed palatability predictors show little to no variation because they they were sampled from the same experiment with a relatively homogeneous set of animals (e.g., all one sex or all grain fed animals) or treatments such as aging period. 
Without any variation in the predictors, the effect cannot be estimated in a standard model despite their established importance according to accepted meat science principles. Naively estimating muscle$\times$cook predictive models without due consideration for the underlying challenges could result in erratic prediction accuracies.
The current MSA model carefully incorporates industry and scientific knowledge in a bespoke manual process to ensure sensible predictions given the challenging data. Our goal is to derive a more automated approach.

In this paper, we introduce a predictive method that bridges a spectrum of muscle$\times$cook-specific models to tackle these challenging data specificities. 
At one extreme, each muscle$\times$cook combination has its own, independent predictive model. 
At the other extreme, a pooled predictive model is obtained across all muscle$\times$cook combinations. 
While neither of the extremes is likely to be desirable in practice, they cover an interesting spectrum of muscle$\times$cook-specific candidate models. 
To estimate this spectrum, we rely on the fused-lasso \citep{Tibshirani05} which encourages similar coefficient estimates across muscle$\times$cook combinations.

Fused lasso-based predictive models show good performance across
various application domains, ranging from
gene regulatory networks in genomics
(e.g., \citealp{omranian2016gene}), and
brain regions from fMRI in neuroscience (e.g., \citealp{zille2017fused}), over
cross-category management in marketing (e.g., \citealp{wilms2018multiclass}), to
price dynamics in commodity markets (e.g., \citealp{barbaglia2016commodity}), amongst others.
To the best of our knowledge, we are the first to devise a regularization-based predictive model tailored towards muscle$\times$cook-specific beef eating quality predictions. 
In doing so, we pair modern data-driven statistical methodology with domain specific expertise following discussions with industry leaders. For the meat science industry, our method provides data driven insights into the similarities between muscle$\times$cook combinations and enables domain experts to interrogate the spectrum of estimated models. From a data science perspective, we tailor the fused-lasso methodology for usage on unbalanced data (varying availability of samples and predictors across groups) with low-signal-to-noise ratio.

We empirically test our predictive model on two unique MSA data sets provided to us thanks to a research collaboration.
First, the proposed regularized predictive model delivers a visual display of so called coefficients paths that provide interesting insights into whether effects of palatability predictors such as carcase weight on beef eating quality are shared across muscle$\times$cook combinations, and if so
among which ones.
Second, we compare our model's predictive performance against several industry benchmarks and find that important gains in prediction accuracy are obtained.
Moreover, across a wide range of muscle$\times$cook combinations, we find our method to offer the most stable prediction performance, thereby ranking among the best both for muscle$\times$cook combinations with many observations as well as few.

The remainder of the paper is organized as follows.
Section \ref{sec_MSA} reviews the MSA grading system for beef palatability and discusses its data specificities.
Section \ref{sec_model} introduces the muscle$\times$cook-specific predictive modeling approach.
Section \ref{sec_MSAresults} presents the results on the muscle$\times$cook-specific predictive models, and Section \ref{sec_MSA_predictions} assesses their predictive performance. Finally, Section \ref{sec_conclude} concludes.

\section{MSA Grading System for Beef Palatability} \label{sec_MSA}

The MSA grading system for beef palatability has been extensively investigated (Section \ref{literature_MSA}). 
It is underpinned by strict protocols for consumer testing, well defined carcase grading methods and muscle categorisation such that a beef eating quality score can be estimated from key palatability predictors in accordance with accepted meat science principles (Section \ref{data_MSA}). 

\subsection{Literature Review} \label{literature_MSA} 

Meat Standards Australia (MSA) is an initiative of the industry body Meat and Livestock Australia to improve the eating quality consistency of beef. 
As part of its mandate, MSA has funded many experiments over the past 20 years, including taste tests of more 1.2 million samples by 170,000 consumers mostly from Australia but also from France, Poland, UK, USA, Japan, South Korea and South Africa.
The collected data serves a number of purposes. 
Individual experiments typically focus on particular questions of interest for the red meat industry, for example packaging effects \citep{polkinghorne2018packaging} or stress \citep{loudon2019stress}. 
When combined across all experiments, the data is used to underpin the MSA beef eating quality prediction model which is implemented in abattoirs across Australia. From July 2020 to June 2021, more than half of the total cattle slaughter were submitted for MSA grading \citep{beqi2021}.

The MSA model is internationally regarded as the leading total quality management (TQM) grading approach to predict beef palatability \citep{bonny2018update}. 
To this end, it combines  key features such as animal-specific traits, process control in several sectors of the beef chain, muscle-specific quality differences and a focus on customer preferences \citep{tatum2006pre}. 
Since consumers are willing to pay for increments in eating quality \citep{polkinghorne2008}, a consumer-oriented grading system with payments based on consumer satisfaction provides a powerful tool to transmit economic incentives through the entire beef supply chain to raise quality consciousness, stimulate industry change, and share premiums throughout the industry (e.g., \citealp{polkinghorne2010meat, smith2008international}).

The Australian beef industry has directly benefited from the implementation of the MSA model through a reduction in the decline of beef consumption and substantial premiums for retailers, wholesalers and producers (e.g., \citealp{bonny2018update, polkinghorne2018commodity, polkinghorne2010meat}). 
Furthermore, its effectiveness in predicting beef palatability has been proven internationally (in France , Ireland, Japan, New-Zealand, Northern Ireland, Poland, South Africa, South Korea, and the USA, see \citealp{bonny2018update} and references therein).

A unique aspect of the consumer-oriented MSA grading system concerns
the choice of a cooked meal portion as its base unit \citep{polkinghorne2010meat}. 
Grades are not assigned to entire carcasses but to \textit{muscles} being aged for a defined number of days and prepared by a particular \textit{cooking method}.
Single carcass grades, as produced by most grading systems, are not able to accurately reflect palatability when being produced from various production systems \citep{watson2008development}. Such grades generally account for less than 15\% of the variance in consumer taste panel scores \citep{bonny2018update, polkinghorne2010meat}. 
Instead, muscle$\times$cook-specific grades better reflect the influence of muscle aging and alternative cooking methods on the differences in expected eating quality performance. 

Reliable quantitative data on the product acceptability of such individual muscle$\times$cook combinations are obtained through  extensive consumer taste panels. 
Such liking or preference testing forms one of the most well-established methods in sensory and consumer evaluation \citep{torrico2018novel}.
Untrained consumers each eat samples (muscles) cooked to a certain protocol and evaluate four sensory traits: tenderness, juiciness, flavour liking and overall liking using line scales. These evaluations are then converted into numerical scores between 0 and 100. A powerful feature of the MSA database is its robustness in number of consumers: over 100,000 consumers have participated in MSA protocols for testing over 700,000 beef samples \citep{torrico2018novel}.

To obtain beef eating quality predictions, a multiple regression approach is used. 
Various  predictors of palatability from the production, processing and value adding sectors are included in the model to predict a carefully constructed meat quality score, \texttt{MQ4}, of individual muscles for a range of cooking methods. 
Despite its simplicity and successful development, the current predictive modeling approach has some potential shortcomings.
First and foremost, less accurate predictions may be obtained for rare muscle-cook combinations since the corresponding regression model either contains few observations or principal palatability predictors can not be included due to data missingness or insufficient variation in the predictor values. To counteract this, often a partial pooling approach is undertaken where similar muscle-cook combinations are manually pooled and combined coefficients are estimated.
While this approach is bespoke, it is also somewhat ad hoc and is certainly labor intensive to update and refine. The proposed regularization-based predictive modeling approach facilitates partial pooling in a data-driven way. This allows us to efficiently find sets of muscle$\times$cook combinations that share estimated coefficients. Experts views can be easily incorporated to inform the pooling.

In the following subsection, we discuss the meat quality score, palatability predictors and challenging data features in more detail.

\subsection{Data} \label{data_MSA}

\paragraph{Beef Eating Quality} 
The MSA model constructs on a single palatability or meat quality score, called \texttt{MQ4}, from the consumer panel tests. 
These sensory experiments source untrained consumers from different community groups and each sample of a muscle-cook combination is tasted by ten participants.
Apart from the two highest and lowest scores which are clipped, an average of the remaining six scores is used in the construction of the quality score.
These average scores on tenderness, juiciness, flavour liking and overall liking are then weighted by 0.3, 0.1, 0.3 and 0.3 respectively to obtain the composite \texttt{MQ4} score.
The weights have been statistically derived from a discriminant analysis \citep{watson2008consumer}.
Predicting overall beef eating quality is deemed equivalent to predicting the \texttt{MQ4} score.

\paragraph{Palatability Predictors} 
When carcases are graded under the MSA model, a variety of traits are recorded, including carcase weight (Kg), sex, hump height (mm), rib fat depth (mm), ossification, MSA marbling score. A declaration is also provided by the producer specifying the feed type (e.g., grass fed or grain fed) and whether or not the animals are a tropical breed (Bos indicus). These carcase level predictors are combined with an estimate of the number of days that the product is expected to be aged before consumption as an input into a muscle level eating quality prediction for each cooking method. \cite{watson2008development} and \cite{bonny2018update} provide a overview on the expected effects of each of these standard traits in classical modelling situations.

\paragraph{Data Specificities}
To illustrate our proposed method, we obtained a sample of eating quality data from Meat and Livestock Australia. The data consists of samples taken on 17 different muscles across three cooking methods. 
For the purposes of this paper, we have split the sample such that we analyse data set ``A'', consisting of a single cooking method (roast) across 14 muscles, separately to data set ``B'', which contains has two similar but distinct slow cooking methods across 13 muscles. 
When we reference the classes defined by combinations of muscles and cooks in data set A, we refer to them as A1 through A14 and similarly for data set B we have classes B1 through B13.

Table \ref{table1} presents the descriptive statistics for our dependent variable and predictors across three cooking methods, roast (RST) and two slow cooking methods (SC1 and SC2). 
While in aggregate there is good range in the predictors, this is not necessarily the case for each cooking method within each muscle type.

\begin{table}
\centering
\caption{Descriptive statistics for the response, eating quality (MQ4) score, and predictors broken down by cooking method (RST, SC1 and SC2).} \label{table1}
\small
\begin{tabular}[t]{llll}
\toprule
Response or Predictor  & RST & SC1 & SC2\\

 & ($n=2031$) & ($n=368$) & ($n=838$)\\ \midrule
\addlinespace[0.3em]
\multicolumn{1}{l}{\textbf{MQ4 score}} & &&  \\
\hspace{1em}Mean (SD) & 55.6 (15.0) & 43.3 (13.8) & 53.6 (15.5)\\
\hspace{1em}Median [Min, Max] & 56.6 [10.6, 95.3] & 42.8 [9.00, 82.2] & 55.6 [5.67, 95.7]\\
\addlinespace[0.3em]
\multicolumn{1}{l}{\textbf{Days aged - dagd } } & &&  \\
\hspace{1em}Mean (SD) & 14.1 (6.43) & 19.1 (6.95) & 13.0 (8.96)\\
\hspace{1em}Median [Min, Max] & 14.0 [3.00, 35.0] & 19.0 [6.00, 35.0] & 8.00 [4.00, 28.0]\\
\hspace{1em}Missing & 0 (0\%) & 122 (33.2\%) & 0 (0\%)\\
\addlinespace[0.3em]
\multicolumn{1}{l}{\textbf{Carcase weight (kg) - cwt}}& &&  \\
\hspace{1em}Mean (SD) & 284 (75.8) & 356 (88.6) & 320 (64.9)\\
\hspace{1em}Median [Min, Max] & 275 [163, 613] & 325 [191, 517] & 321 [171, 468]\\
\hspace{1em}Missing & 42 (2.1\%) & 77 (20.9\%) & 21 (2.5\%)\\
\addlinespace[0.3em]
\multicolumn{1}{l}{\textbf{Hump height (mm) - hump}}& &&  \\
\hspace{1em}Mean (SD) & 68.8 (23.5) & 76.3 (17.7) & 81.9 (30.6)\\
\hspace{1em}Median [Min, Max] & 70.0 [25.0, 210] & 80.0 [40.0, 120] & 75.0 [35.0, 210]\\
\hspace{1em}Missing & 0 (0\%) & 27 (7.3\%) & 0 (0\%)\\
\addlinespace[0.3em]
\multicolumn{1}{l}{\textbf{Rib fat depth (mm) - rbf}}& &&  \\
\hspace{1em}Mean (SD) & 7.81 (4.45) & 9.79 (5.98) & 9.34 (4.26)\\
\hspace{1em}Median [Min, Max] & 7.00 [1.00, 31.0] & 9.00 [2.00, 24.0] & 8.50 [3.00, 24.0]\\
\addlinespace[0.3em]
\multicolumn{1}{l}{\textbf{Ossification score - uoss}}& &&  \\
\hspace{1em}Mean (SD) & 187 (105) & 188 (44.4) & 150 (32.8)\\
\hspace{1em}Median [Min, Max] & 150 [100, 590] & 170 [110, 270] & 140 [100, 270]\\
\hspace{1em}Missing & 132 (6.5\%) & 77 (20.9\%) & 192 (22.9\%)\\
\addlinespace[0.3em]
\multicolumn{1}{l}{\textbf{MSA marbling score - umb}}& &&  \\
\hspace{1em}Mean (SD) & 308 (159) & 357 (165) & 331 (125)\\
\hspace{1em}Median [Min, Max] & 280 [130, 1180] & 310 [130, 810] & 320 [150, 810]\\
\addlinespace[0.3em]
\multicolumn{2}{l}{\textbf{Tropical breed content - bi}}& & \\
\hspace{1em}No TBC & 1356 (66.8\%) & 141 (38.3\%) & 531 (63.4\%)\\
\hspace{1em}TBC & 609 (30.0\%) & 156 (42.4\%) & 129 (15.4\%)\\
\hspace{1em}Missing & 66 (3.2\%) & 71 (19.3\%) & 178 (21.2\%)\\
\addlinespace[0.3em]
\multicolumn{1}{l}{\textbf{Feed type - feed}}& &&  \\
\hspace{1em}Grain & 1260 (62.0\%) & 75 (20.4\%) & 537 (64.1\%)\\
\hspace{1em}Grass & 705 (34.7\%) & 100 (27.2\%) & 200 (23.9\%)\\
\hspace{1em}Missing & 66 (3.2\%) & 193 (52.4\%) & 101 (12.1\%)\\
\addlinespace[0.3em]
\multicolumn{1}{l}{\textbf{Sex - sex}} & &&  \\
\hspace{1em}Female & 583 (28.7\%) & 0 (0\%) & 25 (3.0\%)\\
\hspace{1em}Male & 857 (42.2\%) & 0 (0\%) & 103 (12.3\%)\\
\hspace{1em}Missing & 591 (29.1\%) & 368 (100\%) & 710 (84.7\%)\\
\bottomrule
\end{tabular}

\end{table}

\begin{figure}[t]
    \centering
    \includegraphics[width = 0.45\textwidth]{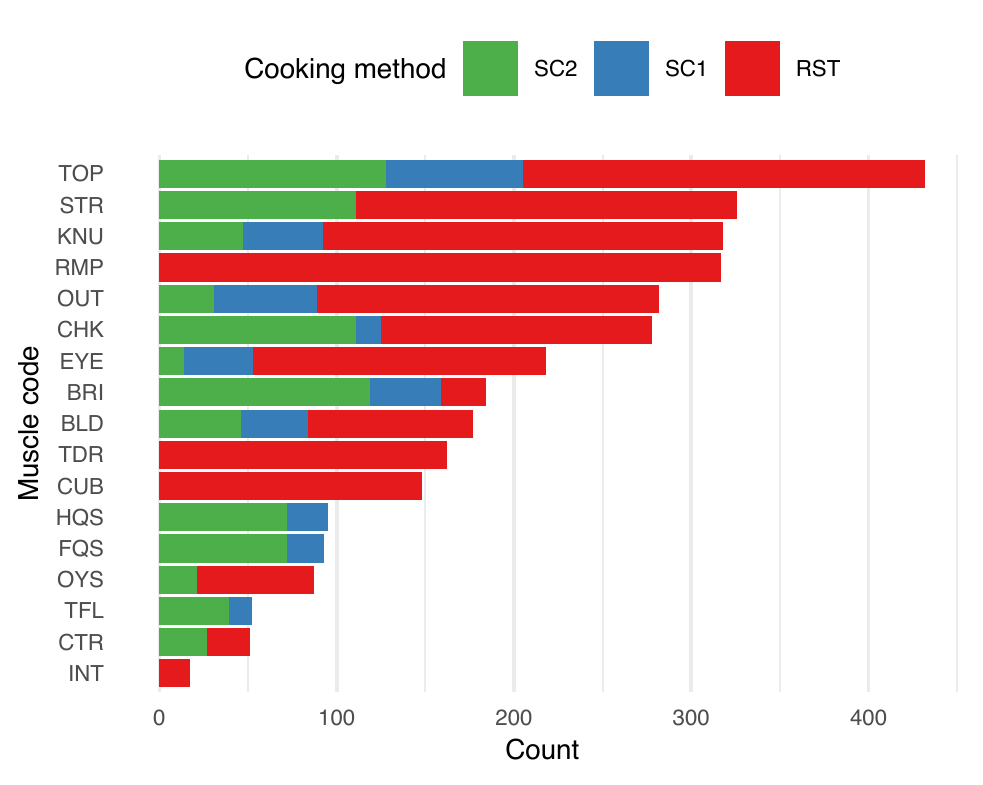}
    \caption{
For each muscle (row), we display the number of observed samples on the three cooking methods (SC2 in green, SC1 in blue, RST in red).
    }
    \label{fig:unbalanced}

\end{figure}

One of the key challenges we face when working with this data is the unbalanced nature of the classes (i.e., muscle$\times$cook combinations), as highlighted in Figure \ref{fig:unbalanced}. We see some muscles only have one cooking method or very limited samples, whereas other muscles have quite substantial numbers of observations across all three cooking methods.

\begin{figure}[t]
    \centering
    \includegraphics[width = 0.45\textwidth]{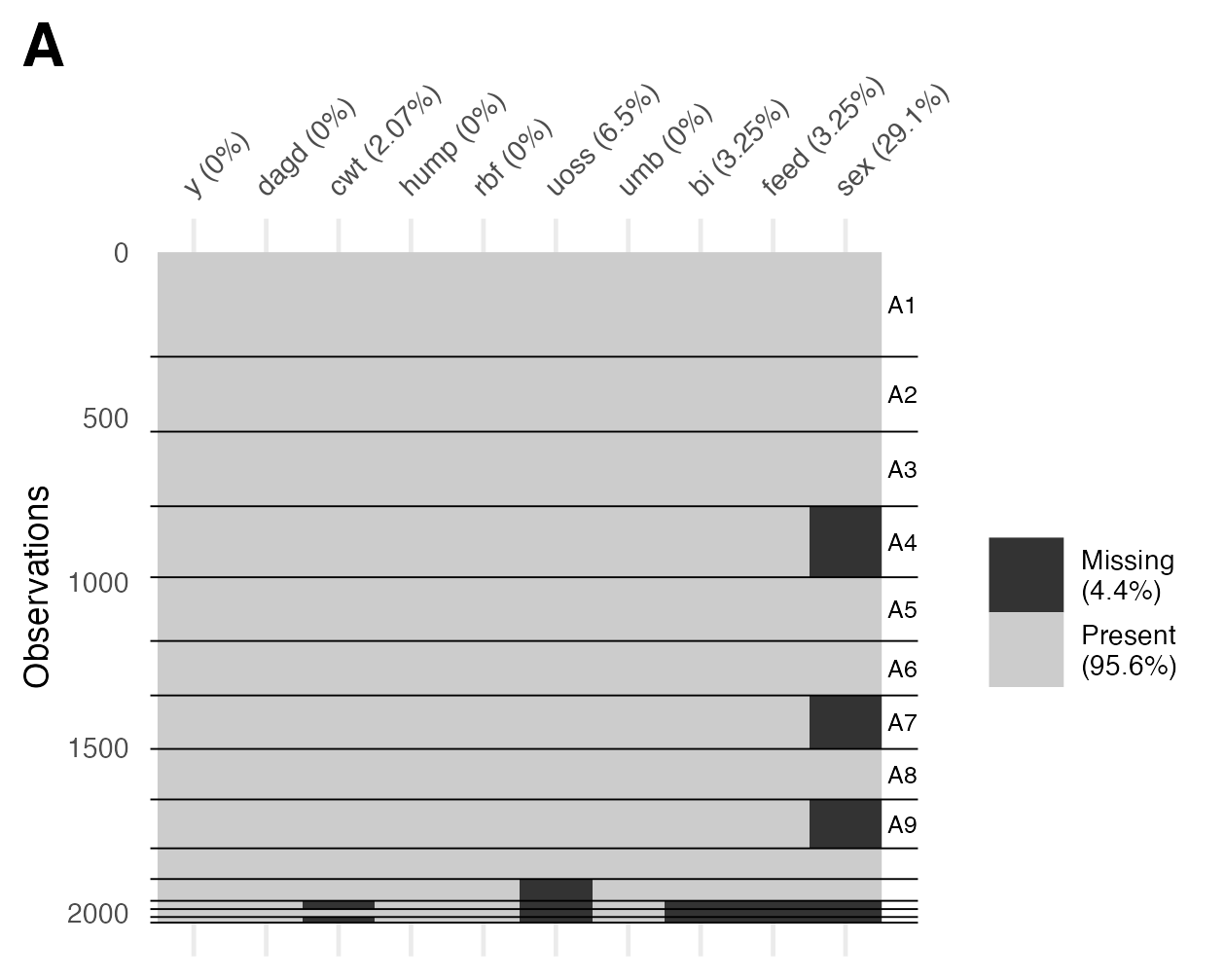}
    \includegraphics[width = 0.45\textwidth]{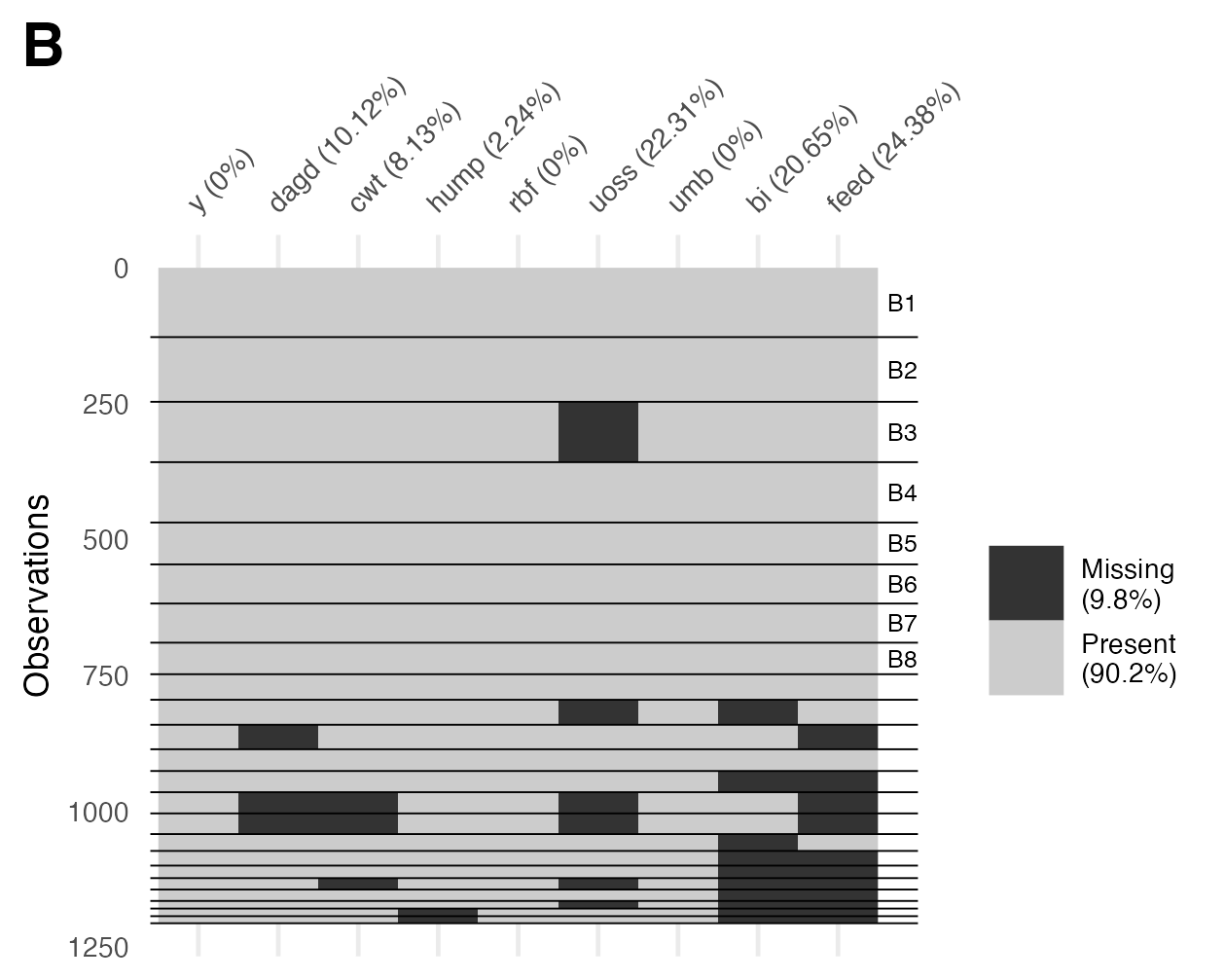}
    \caption{Missingness for the two data sets. 
    The horizontal lines define boundaries among classes  (i.e.,\ muscle$\times$cook combinations) and the first few classes in each have been enumerated down the right hand side of the plot.} 
    \label{fig:missingness}
\end{figure}

Imbalanced class sizes is only a first challenging aspect of this data, data missingness is a second. The second data set has a greater number of muscles with smaller sample sizes and more missing data than the first data set. Figure \ref{fig:missingness} presents an overview of the two data sets and the missingness in each using a heat map \citep{visdat}. 
The observations are sorted by class with the largest class at the top of the figure and smallest classes down the bottom. 
The first few large classes are named sequentially down the right hand side of each panel. 

In data set A (left panel) we have larger and fewer classes than in data set B (right panel) that has more classes which tend to be smaller with more missingness.
Missingness may arise systematically, for example in older experiments where a particular trait was not regularly measured 
or randomly through input errors identified in subsequent data validation routines.
The chunks of missingness have been defined such that if there was a missing observation on an animal in a muscle$\times$cook combination, then all animals were assigned missing values for that trait. 
The drawback of this approach is that for these muscle$\times$cook combinations, if we were to estimate models for each combination separately, the coefficients for the impacted variables would be unidentifiable. 
Further, for predictive models that pool across muscle$\times$cook combinations, we would need to exclude any trait with missingness to ensure we retain all the observations. 
In the current example, we see missing values in important traits such as feed type, ossification score and days aged.

\section{A ``Muscle$\times$Cook"-Specific Predictive Model}  \label{sec_model}
We use a linear regression model to predict beef eating quality for muscle$\times$cook specific combinations within the MSA framework. 
A linear model is a standard approach to modelling eating quality data derived from experiments that follow the MSA protocols due to its simplicity and interpretability.
Let ${y}^{(m)}_i$ denote the response or dependent variable being the measure of eating quality for sample (i.e.\ observation) $i=1, \ldots, n_m$ belonging to muscle$\times$cook combination $m=1, \ldots, M$. 
Denote the $p_m$ palatability predictor variables for muscle$\times$cook combination $m$ by $\{x_{ij}^{(m)}\}_{1 \leq j \leq p_m}$. 
Note that some values on the palatability predictors might be missing for some muscle$\times$cook combinations. 
We denote the set of available predictors for muscle$\times$cook combination $m$ by $\mathcal{S}_m$ with cardinality $p_m$.
We thus allow the number of observations $n_m$ and the number of predictors $p_m$ to vary across the muscle$\times$cook combinations.

It is common practice to estimate a separate regression model 
\begin{equation}
{y}^{(m)}_i = \beta_0^{(m)} + \beta_1^{(m)}{x}^{(m)}_{i1} + \beta_2^{(m)}{x}^{(m)}_{i2} + \ldots  + \beta_{p_m}^{(m)}{x}^{(m)}_{ip_m} + \varepsilon_{i}^{(m)}, \  \ \ i = 1, \ldots, n_m,
\end{equation}
for each muscle$\times$cook combination $m$. Here, $\beta_0^{(m)}$ is the intercept, $\beta_1^{(m)}, \ldots,\beta_{p_m}^{(m)}$  are unknown regression parameters,  and $\varepsilon_{i}^{(m)}$ are the error terms for muscle$\times$cook combination $m=1, \ldots, M$.
Traditionally, parameter estimates are obtained  for each muscle$\times$cook combination separately by running $m$ independent least squares regressions. 
We, instead, propose to \textit{jointly} estimate the $m$ regression models such that strength across muscle$\times$cook combinations can be borrowed in the estimation process. To this end, we use a regularization-driven predictive model.

\subsection{Predictive Model and Estimator}
To obtain beef eating quality predictions for all muscle$\times$cook combinations, we consider the joint predictive model 
\begin{equation}
y_i^{(m)} = {{\bf x}_{i}^{(m)}}^\top \boldsymbol \beta^{(m)} + \varepsilon_i^{(m)}, \label{multiclassreg}
\end{equation}
for $m=1\ldots, M$, where ${\bf x}_{i}^{(m)} = (x_{i1}^{(m)}, \ldots, {x}^{(m)}_{ip})^\top$, ${\boldsymbol \beta}^{(m)} = (\beta_1^{(m)}, \ldots,\beta_{p}^{(m)})^\top$, $p$ is the total number of available predictors and  ${\bf u}^\top$ denotes the transpose of a vector ${\bf u}$.
If the $j$th predictor is not available for a particular muscle$\times$cook combination it does not enter the model and the corresponding coefficient, $\beta_j^{(m)}$, is not estimated.
Denote the joint regression parameter vector by $\boldsymbol \beta = ({\boldsymbol \beta^{(1)}}^\top, \ldots, {\boldsymbol \beta^{(M)}}^\top)^\top$. 
We define its estimator as 
\begin{equation}
\widehat{\boldsymbol \beta}_\lambda = \underset{\boldsymbol \beta}{\operatorname{argmin}} \ \dfrac{1}{2} \sum_{m=1}^{M} \sum_{i=1}^{n_m} (y_i^{(m)} - {{\bf x}_{i}^{(m)}}^\top \boldsymbol \beta^{(m)})^2 + \lambda \sum_{m< m^\prime}^{M} \sum_{j=1}^{p} w_j^{(m,m^\prime)}| \beta_j^{(m)} - \beta_j^{(m^\prime)}|. \label{fused}
\end{equation}
The first term in \eqref{fused} is a least squares criterion. 
The second term is a fusion penalty (see e.g., \citealp{Tibshirani05}) on the parameters with tuning parameter $\lambda>0$ and weights $w_j^{(m,m^\prime)}$ for each pair of muscle$\times$cook combinations $m$ and $m^\prime$. We take a fusion penalty to induce similarity in corresponding coefficients across muscle$\times$cook combinations. 

The larger the value of the tuning parameter $\lambda$, the more corresponding elements of $\widehat{\boldsymbol \beta}^{(1)}, \ldots, \widehat{\boldsymbol \beta}^{(M)}$ will be, hence shared, across muscle$\times$cook combinations. 
If $\lambda \rightarrow \infty$, all regression coefficients on the available corresponding predictors across muscle$\times$cook combinations will be estimated identically. 
Then $\widehat{\boldsymbol \beta}$ corresponds to a \textit{pooled least squares} estimator. 
In the remainder, we refer to this pooled estimator as the ``new pooled" estimator since it pools effects of  corresponding palatability predictors across those muscle$\times$cook combinations for which the palatability predictor is available, and this in contrast to the ``classic pooled" estimator which requires the same set of palatability predictors across muscle$\times$cook combinations to pool their effects.
If $\lambda = 0$, then each muscle$\times$cook combination $m$ has its own, independent predictive model and there are no similarities across muscle$\times$cook combinations.
Then $\widehat{\boldsymbol \beta}$ corresponds to the \textit{separate least squares} estimator. 
By varying the value of $\lambda$ a \textit{spectrum} of muscle$\times$cook-specific predictive models is thus obtained.

In our eating quality application, the number of observations is highly unbalanced across muscle$\times$cook combinations.
Muscle$\times$cook combinations with only a few observations should be encouraged to  borrow strength from muscle$\times$cook combinations with many observations to improve estimation and prediction accuracy. 
We therefore use the following weights in equation \eqref{fused},
\begin{equation} \label{normalize_weight}
w_j^{(m,m^\prime)} =
\begin{cases}
\widetilde{w}^{(m,m^\prime)}/\underset{(m,m^\prime)}{\text{max}}\widetilde{w}^{(m,m^\prime)} & \text{if predictor} \: j \in \mathcal{S}_m \ \text{and predictor} \ j \in \mathcal{S}_{m^\prime} \\
0 & \text{otherwise},
\end{cases}
\end{equation}
with
$\widetilde{w}^{(m,m^\prime)} =\text{max}(n_m,n_m^\prime)/\text{min}(n_m,n_m^\prime)$
the ratio of the largest sample size  to the smallest.
This way, unbalanced muscle$\times$cook combinations (i.e.,\ having unequal sample sizes) are encouraged to be estimated similarly before balanced muscle$\times$cook combinations (i.e.,\ having equal sample sizes). 
For unbalanced  combinations, the regression coefficients of the rare combinations (having few samples) will be pulled towards those of the more common combinations (having many samples).

\subsection{Algorithm \label{algorithm}} 

As common in the regularization literature, the predictors are standardized before estimating the joint predictive model.
To solve optimization problem \eqref{fused}, note that it can be re-written as a generalized lasso problem \citep{Tibshirani11},
\begin{equation}
\widehat{\boldsymbol \beta}_\lambda = \underset{\boldsymbol \beta}{\operatorname{argmin}} \ \dfrac{1}{2} \sum_{m=1}^{M} \sum_{i=1}^{n_m} (y_i^{(m)} - {{\bf x}_{i}^{(m)}}^\top \boldsymbol \beta^{(m)})^2 + \lambda ||{\bf D}\boldsymbol\beta||_1, \label{generalizedlasso}
\end{equation}
where ${\bf D}$ 
is a predefined matrix capturing the coupling of coefficients across muscle$\times$cook combinations.
Its number of columns is equal to the dimension of the parameter vector $\boldsymbol{\beta}$, 
its number of rows is 
equal to the total number of pairs of corresponding regression parameters  across  muscle$\times$cook combinations. Each row contains two non-zero entries, $w_j^{(m,m^\prime)}$ and $-w_j^{(m,m^\prime)}$, allowing for the coupling between a shared predictor $j$ of muscle$\times$cook combinations $m$ and $m^\prime$.

Optimization problem \eqref{generalizedlasso} in generalized fused lasso form for fixed $\lambda$ can be solved through the path algorithm of 
\cite{Tibshirani11}, implemented in the \verb|R|-package \verb|genlasso| \citep{genlasso}. 
All computations are carried out in \verb|R| version 4.1.3  \citep{r2021}. The code of the algorithm is available on the GitHub page of the first author.

Each choice of the tuning parameter $\lambda$ thus corresponds to a muscle$\times$cook-specific model. 
By solving problem \eqref{generalizedlasso} for a grid of $\lambda$-values, we obtain an entire spectrum of muscle$\times$cook-specific models. 
To this end, we take a grid of $\lambda$-values on log-linear scale in the interval [0, $\widehat{\lambda}_{\text{max}}$]. 
Here, $\lambda = 0$ corresponds to separate predictive models for each muscle$\times$cook combination and $\lambda = \widehat{\lambda}_{\text{max}}$ corresponds to the ``new pooled" predictive model.
In the following sections, we discuss estimation and prediction performance across this spectrum of muscle$\times$cook-specific models.

\section{Modeling MSA Muscle$\times$Cook-Specific Beef Eating Quality} \label{sec_MSAresults}

\begin{figure}
    \centering
    \includegraphics[width = 0.95\textwidth]{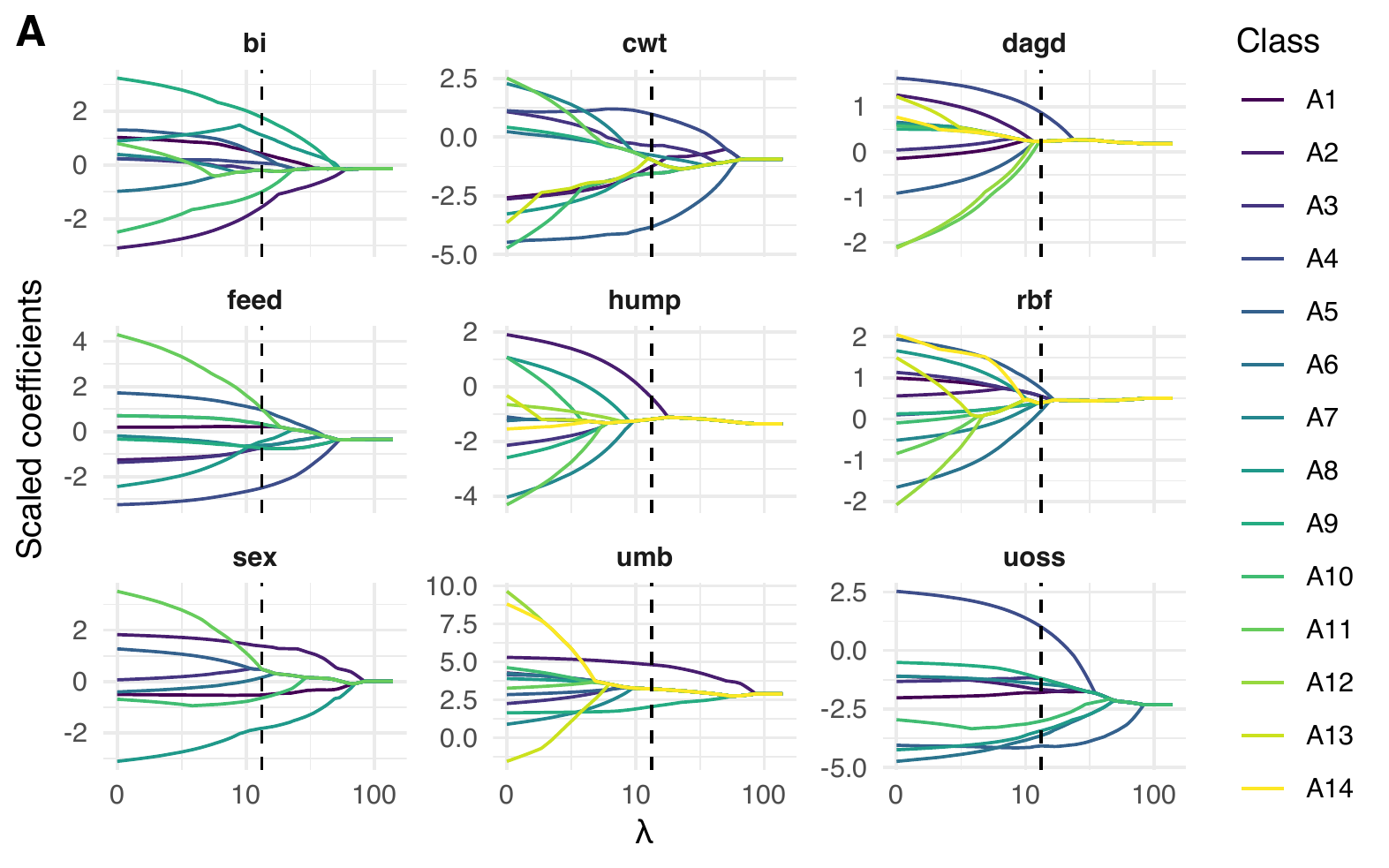} \\
    \includegraphics[width = 0.95\textwidth]{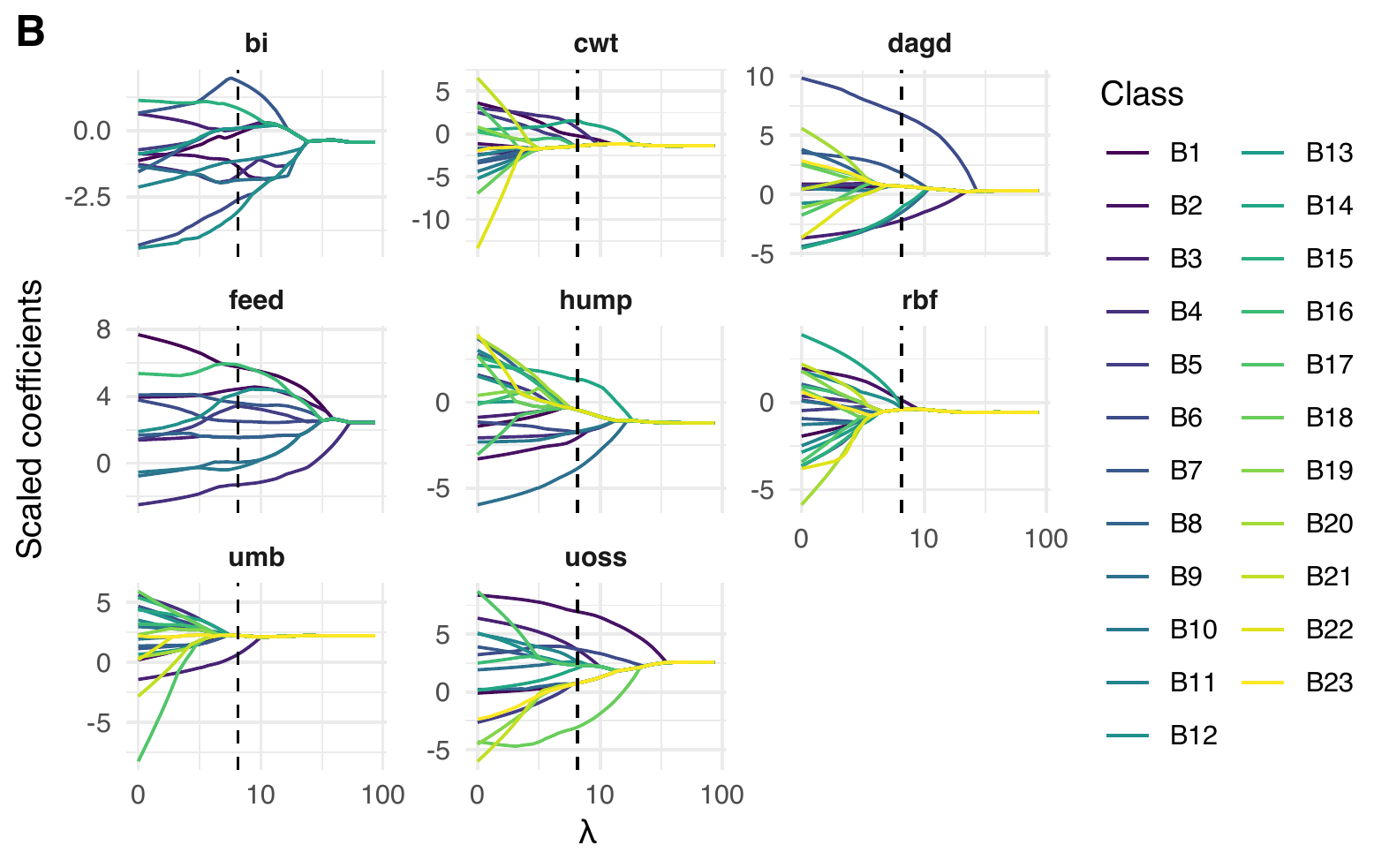}
    \caption{Coefficient paths for data set A (top) and data set B (bottom). Each line represents the shrinkage of a coefficient for one muscle$\times$cook combination (i.e. class) along the regularization path. The classes are ordered from the largest to smallest number of observations. The dashed vertical line indicates the value of the tuning parameter picked by the AIC. 
    }
    \label{fig:coefficientpaths}
\end{figure}

For each of the two data sets introduced in Section \ref{data_MSA}, we obtain the parameter estimates in the multiple regression model \eqref{multiclassreg} through the fused lasso estimator. 
Figure \ref{fig:coefficientpaths} shows the estimated scaled coefficient paths across the full range of tuning parameter values for both data sets. 
At $\lambda=0$, we  obtain the separate least squares model, as $\lambda$ increases on the horizontal axis, the coefficients tend to start being pooling together until we end up at the ``new" pooled least squares solution (i.e.,\ single line for each predictor effect). 

These coefficient plots help guide further refinement of the eating quality model in a number of ways. 
Firstly, some coefficient paths join together at low levels of the tuning parameter, such as for the predictors MSA marbling score (umb) or rib fat depth (rbf)  in panel B of Figure \ref{fig:coefficientpaths}.
Here, a small amount of regularisation leads to pooling of the majority of muscle$\times$cook combinations.
This suggest similarity between the response for those muscle$\times$cook combinations to changes in specific measured traits and these predictors could reasonably be modelled with a common coefficient across the majority of muscle$\times$cook combinations. 
In contrast, some coefficient paths remain stubbornly separate from the pooled path, such as for feed type (feed) in data set B. 
Such cases indicate muscle$\times$cook combinations that do not necessarily conform to standard, shared responses. 

Secondly, we can visualise the range and diversity of estimated coefficients across muscle$\times$cook combinations. 
It is often the case that there is some initial rapid shrinkage of a coefficient followed by a more gradual joining of paths. 
In Figure \ref{fig:coefficientpaths} the muscle$\times$cook combinations (i.e.\ classes) are ordered from largest sample size to smallest sample size. 
The relative sizes of each class can be seen in Figure \ref{fig:missingness}. 
The darker purple lines represent larger muscle$\times$cook combinations whereas the lighter yellow lines have smaller samples sizes.
Where we have missing observations in a class for a given predictor we do not observe a coefficient path.
In panel A we often see the smaller groups (lighter lines) rapidly shrinking towards other coefficient paths. 
This is particularly notable in the carcase weight (cwt), hump height (hump) and rib fat depth (rbf) predictors. There, the smaller groups join with others at small values of the tuning parameter $\lambda$.
This confirms our prior belief that the separate least squares solution is likely much too noisy, particularly for classes with smaller sample sizes and even a small amount of regularisation is very beneficial for its stabilising influence.

Finally, these plots provide insight into which muscle$\times$cook combinations may require further individual investigation. 
For example, there are occasionally some very stubborn coefficient paths in large groups that behave differently to the other groups. 
An example of this is group A4 which has quite distinct paths for ossification (uoss) and days aged (dagd) and feed type (feed).

The regularization paths thus provide a useful visual tool to balance flexibility (for small values of $\lambda$) with model parsimony (for large values of $\lambda$) when building an effective eating quality model. 
These plots can be combined with prior scientific understanding about  how different muscles should respond to changes in underlying traits,
to arrive at a good choice for the tuning parameter $\lambda$, thereby selecting an ``optimal" model along this spectrum. 
In particular, an experienced researcher may be able to identify where the regularisation has gone too far and the resulting estimated coefficients have been too heavily shrunk. 
Indeed, we often see this when trying to include too many disparate classes in a single model. For example if all classes across data sets A and B were combined in a single model the algorithm struggles to extract meaningful structure from the data. This reflects a real world understanding that the roast cooking method (data set A) has meaningfully different response characteristics to the slow cook method (data set B) and it is not necessarily sensible to try to pool these together in a combined analysis.

The real power of the methodology is likely to be revealed when domain experts and data scientists work together to identify sensible and biologically meaningful models along the whole spectrum of muscle$\times$cook-specific models. However, a practitioner might also want to consider data-driven selection methods. 
To this end, one can rely on various selection criteria such as cross-validation, or information criteria like the Akaike Information Criterion (AIC). 
As an illustration, we use the latter to select an ``optimal"  model from the  spectrum of muscle$\times$cook-specific models since it is simple and fast to compute. 
In particular, for each $\lambda$-value in the grid one computes the value of
$$\text{AIC}_\lambda = 
\log(\hat{\sigma}^2_\lambda) + 2\operatorname{df}_\lambda/n,
$$
where $\widehat{\sigma}_\lambda^2 = \frac{1}{n} \sum_{m=1}^{M}\sum_{i=1}^{n_m}(y_i^{(m)} - {{\bf x}_{i}^{(m)}}^\top \widehat{\boldsymbol\beta}^{(m)}_\lambda)^2$, with total sample size $n=\sum_{m=1}^{M} n_m$ and $\text{df}_\lambda$ refers to the degrees of freedom which equals the number of distinct regression coefficients (e.g., \citealp{Tibshirani11}). Note that the residual variance $\widehat{\sigma}_\lambda^2$ and degrees of freedom  $\text{df}_\lambda$  depend on $\lambda$ since they result from solving optimization problem \eqref{generalizedlasso} with tuning parameter $\lambda$. The ``optimal"  model along the spectrum then corresponds to the one minimizing $\text{AIC}_\lambda$, as indicated by the vertical dashed lines in Figure \ref{fig:coefficientpaths}.
From the top panel in Figure \ref{fig:coefficientpaths}, for instance, we see that AIC recommends to pool the effects across muscle$\times$cook combinations mainly for the predictors days aged (dagd), hump height (hump) and rib fat depth (rbf), whereas  muscle$\times$cook-specific effects are selected for the predictors tropical breed content (bi) and feed type (feed).

\section{Assessing Predictive Performance} \label{sec_MSA_predictions}

We conduct an out-of-sample prediction exercise to evaluate  prediction performance across the spectrum of  muscle$\times$cook combinations.
We split each of the $M$ data sets randomly into $K = 5$ parts of nearly equal size. Each $k^{th}$ block ($k=1, \ldots, K$) is left out once and used as a test set. The remaining blocks are used as training set to estimate the regression model, yielding estimates $\widehat{\boldsymbol \beta}_{\lambda, k}$ for each $\lambda$-value in the grid.
Corresponding predictions
$ \widehat{y}_{i, k}^{(m)} = {{\bf x}_{i, k}^{(m)}}^\top\widehat{\boldsymbol \beta}_{\lambda, k}^{(m)}$ are obtained
for the $1 \leq i \leq \text{ntest}^{(m)}_k$ observations in the $k^{th}$  test data set of muscle$\times$cook combination $m = 1, \ldots, M$. 
We then measure prediction performance for each $\lambda$-value and block $k$ by computing traditional statistical loss measures (Section \ref{stat_loss}) or industry specific loss measures (Section \ref{star_loss}). 

\subsection{Statistical Loss} \label{stat_loss}
We measure prediction accuracy by computing the Mean Absolute Error for each block, $\text{MAE}_{\lambda, k}$, and the 
overall Mean Absolute Error, $\text{MAE}_\lambda$, averaged over all $K$ blocks,
$$ \text{MAE}_{\lambda, k} = \dfrac{1}{M\cdot\text{ntest}^{(m)}_k}\sum_{m=1}^{M} \sum_{i=1}^{\text{ntest}^{(m)}_k} \left|\widehat{y}_{i, k}^{(m)} - {y}_{i, k}^{(m)} \right| \  \text{and} \  \text{MAE}_\lambda = \dfrac{1}{K} \sum_{k=1}^{K} \text{MAE}_{\lambda, k}.$$
Alternatively, one can evaluate prediction accuracy in terms of the Mean Squared Error (MSE). The MSE and MAE results were similar so in the interest of brevity, only the MAE results are reported.

\begin{figure}
    \centering
    \includegraphics[width = 0.75\textwidth]{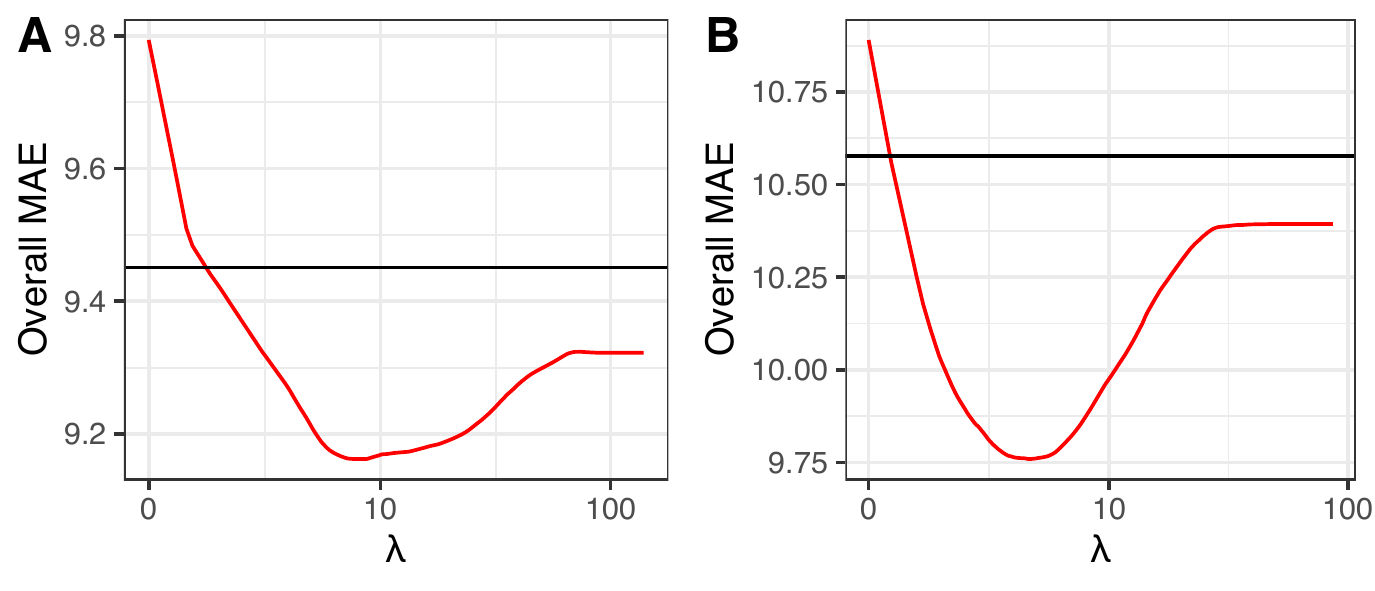}
    \caption{Mean absolute errors for data set A (left) and data set B (right). The black horizontal line represents the corresponding forecast error for the classic pooled least squares model.  
    }
    \label{fig:mafe}
\end{figure}

Figure \ref{fig:mafe} displays the overall MAE for both data sets across the whole spectrum of muscle$\times$cook combinations ranging from the separate least squares model ($\lambda=0$) towards the new pooled least squares model (largest $\lambda$). 
First, we see that best predictive model across the spectrum for model A returns an MAE of 9.16 for data set A and 9.76 for data set B.
Hence, the best models along the spectrum result in important predictive gains over the extremes:
improvements of 6.4\% (10.4\%) compared to the separate least squares for data set A (B); and improvements of 1.7\% (6.1\%) compared to the new pooled least squares.
The later also consistently provides improvements over the classic pooled least squares (i.e.,\ horizontal black line).
Moreover, there is a vast range of tuning parameter values along the spectrum that result in such predictive gains. 
Practitioners can therefore combine these predictive performance path with the coefficients paths from the previous section to trade-off flexibility but generally lower predictive performance (for small values of $\lambda$)  with better predictive performance but model parsimony (for larger values of $\lambda)$.

Next, we zoom into the predictive performance across the different muscle$\times$cook combinations as detailed in Table \ref{tab:mafe_per_class}.
Not only averaged across all muscle$\times$cook combinations, the proposed predictive model performs well, but also when looking at individual combinations. 
The best performing model along the spectrum, labelled ``CV selected" also 
offers the most stable predictive performance, among the best ranked methods across the different different muscle$\times$cook combinations (i.e.,\ classes).
For large classes, its predictive performance is close to the separate least squares solution, which we indeed expect to perform well. 
For small classes, it is competitive and closer to the new pooled least squares solution, since mainly small classes benefit from being pooled with the larger ones. The separate least squares solution looses its competitiveness here since its solutions are too noisy.

\begin{table}
    \caption{Mean absolute  errors per muscle$\times$cook combination (i.e.,\ class) for each of the four predictive methods: CV selected, new pooled, classic pooled and separate least squares.
    }
\captionsetup[table]{labelformat=empty,skip=1pt}
\begin{longtable}{lrrrrr}
\toprule
Class & Sample size & CV selected & New pooled & Classic pooled & Separate \\ 
\midrule
A1 & 317 & $9.24$ & $9.12$ & $9.33$ & $9.30$ \\ 
A2 & 227 & $9.83$ & $10.27$ & $10.31$ & $9.92$ \\ 
A3 & 226 & $11.11$ & $11.15$ & $11.05$ & $11.31$ \\ 
A4 & 215 & $10.91$ & $11.60$ & $11.37$ & $11.05$ \\ 
A5 & 193 & $10.63$ & $11.17$ & $11.94$ & $10.69$ \\ 
A6 & 165 & $8.70$ & $8.77$ & $9.09$ & $8.65$ \\ 
A7 & 162 & $6.35$ & $6.33$ & $6.26$ & $6.43$ \\ 
A8 & 153 & $9.37$ & $9.40$ & $9.88$ & $9.33$ \\ 
A9 & 148 & $8.99$ & $9.38$ & $9.10$ & $9.09$ \\ 
A10 & 93 & $7.98$ & $7.99$ & $8.37$ & $8.41$ \\ 
A11 & 66 & $6.83$ & $7.00$ & $6.98$ & $7.35$ \\ 
A12 & 25 & $9.13$ & $9.25$ & $9.58$ & $9.24$ \\ 
A13 & 24 & $9.55$ & $9.48$ & $9.40$ & $10.36$ \\ 
A14 & 17 & $9.65$ & $9.59$ & $9.66$ & $15.98$ \\ 
&&&&&\\
B1 & 128 & $10.26$ & $11.10$ & $11.01$ & $10.15$ \\ 
B2 & 119 & $11.55$ & $12.63$ & $13.99$ & $11.72$ \\ 
B3 & 111 & $10.74$ & $10.78$ & $10.90$ & $10.64$ \\ 
B4 & 111 & $8.71$ & $9.28$ & $9.04$ & $8.85$ \\ 
B5 & 77 & $8.21$ & $8.25$ & $7.75$ & $8.42$ \\ 
B6 & 72 & $12.13$ & $14.09$ & $16.14$ & $12.05$ \\ 
B7 & 72 & $9.36$ & $9.80$ & $10.50$ & $9.79$ \\ 
B8 & 58 & $9.14$ & $9.57$ & $8.83$ & $9.94$ \\ 
B9 & 47 & $8.97$ & $10.39$ & $11.08$ & $9.52$ \\ 
B10 & 46 & $7.82$ & $7.91$ & $8.22$ & $8.31$ \\ 
B11 & 45 & $9.04$ & $8.91$ & $9.05$ & $10.69$ \\ 
B12 & 40 & $9.27$ & $9.91$ & $11.11$ & $10.04$ \\ 
B13 & 39 & $9.07$ & $8.98$ & $9.15$ & $10.12$ \\ 
B14 & 39 & $10.33$ & $13.48$ & $11.37$ & $9.53$ \\ 
B15 & 38 & $8.03$ & $8.21$ & $8.19$ & $8.38$ \\ 
B16 & 31 & $8.53$ & $9.29$ & $10.71$ & $8.76$ \\ 
B17 & 27 & $11.01$ & $10.83$ & $11.56$ & $11.77$ \\ 
B18 & 23 & $9.81$ & $12.30$ & $12.13$ & $11.64$ \\ 
B19 & 21 & $10.75$ & $11.25$ & $10.52$ & $12.93$ \\ 
B20 & 21 & $8.86$ & $9.32$ & $9.14$ & $10.27$ \\ 
B21 & 14 & $9.13$ & $8.44$ & $8.52$ & $12.29$ \\ 
B22 & 14 & $16.18$ & $15.82$ & $16.41$ & $20.03$ \\ 
B23 & 13 & $7.57$ & $8.49$ & $7.98$ & $14.63$ \\ 
\bottomrule
\end{longtable}

    \label{tab:mafe_per_class}
\end{table}

In sum, while for a particular muscle$\times$cook combination, the CV selected is thus either competitive to or slightly better than one of the two endpoints of the spectrum,
the advantage of the regularised predictive approach is in obtaining a 
stable set of coefficients across all muscle$\times$cook combinations that is never drastically worse in terms of prediction accuracy than either endpoint.

\clearpage
\subsection{Star Ratings and Accuracy} \label{star_loss}
In practice, the predicted eating quality scores are discretized into one of four categories: unsatisfactory ($2^*$), good everyday ($3^*$), better than everyday ($4^*$) or premium quality ($5^*$). 
As a further method of evaluation, we consider overall accuracy and a consumer focused accuracy assessment where a ``correct'' outcome is considered to be a correct categorisation or an incorrect categorisation that favours the consumer, i.e.,\ an under prediction where the truth is a better quality sample than the model predicts. 
The justification for this alternative accuracy assessment is related to consumer purchasing behaviour. 
\cite{polkinghorne2008} note that one of the common reasons cited for declining consumption of beef in countries around the world is inconsistent eating quality experiences. 
A key tenant of an eating quality prediction model is to ensure a minimum standard at various price points. 
When we can reliably ensure that a consumer gets what they paid for or an experience better than they paid for, they are more likely to be repeat purchasers. 

\begin{figure}
    \centering
    \includegraphics[width = \textwidth]{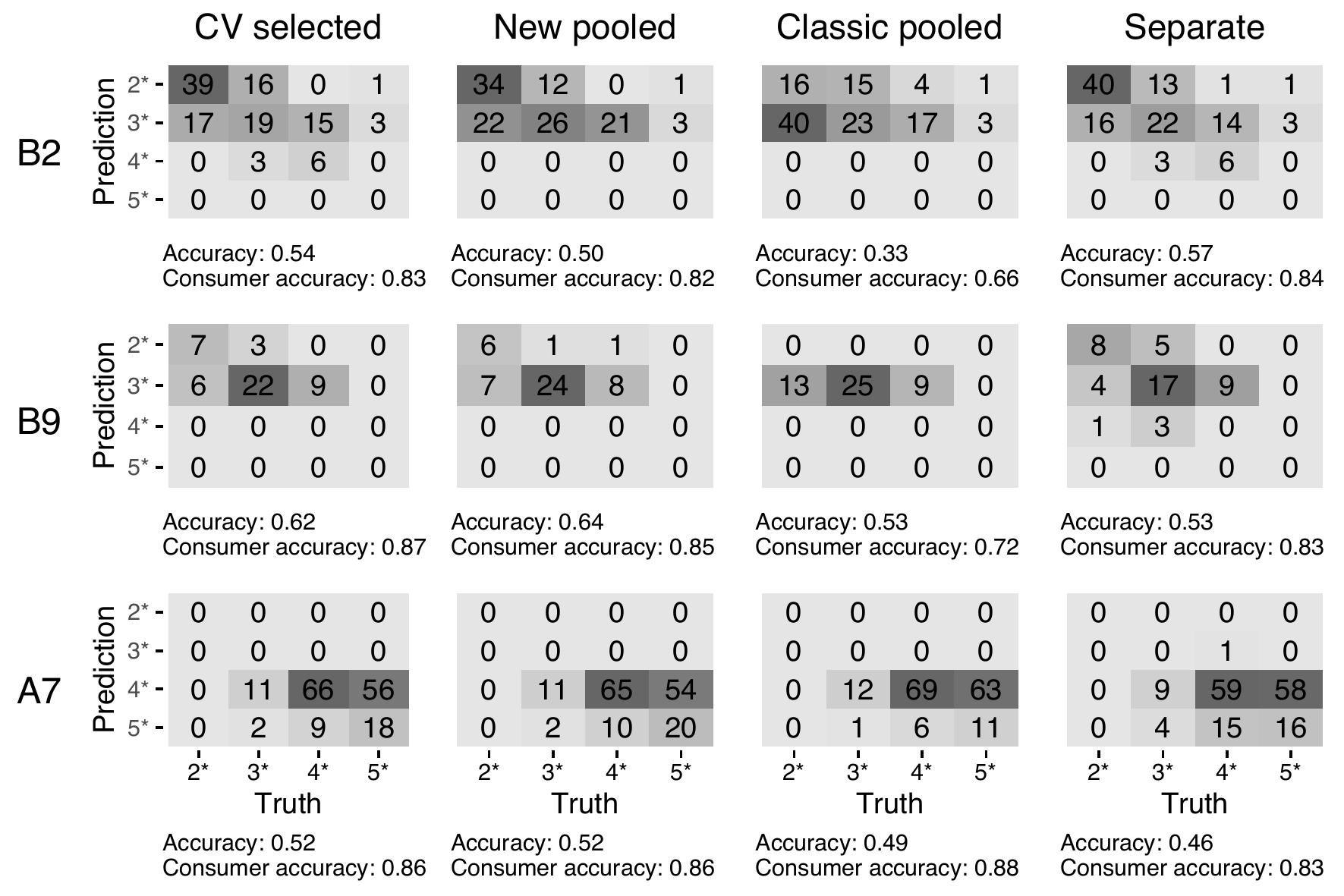}
    \caption{Confusion matrices and accuracy performance outcomes for a selection of muscle$\times$cook combinations.}
    \label{fig:confmat}
\end{figure}

We apply the standard industry thresholds of \cite{watson2008consumer} to categorise the observed eating quality scores and our out-of-sample predicted eating quality scores. 
As expected, there was some variation in the accuracy and consumer focused accuracy across the muscle$\times$cook combinations, particularly in the classes where there are relatively few observations and where there is limited variation in the observed categories. 
Figure \ref{fig:confmat} presents a set of confusion matrices and accuracy measurements for a selection of representative muscle$\times$cook combinations.

We start by discussing the confusion matrices of muscle$\times$cook combination B2 ($n=119$), displayed in the top row of Figure \ref{fig:confmat}, which is a representative example of  muscle$\times$cook combinations with many observations.
For such large classes, we typically see that the proposed method ``CV selected" and the separate least squared method perform the best in terms of the accuracy metrics.
In contrast, the classic pooled method performs poorly.
It has a severe tendency to  over-predict true $2^*$ samples in the $3^*$ category, which in turn negatively impacts the consumer accuracy metrics as consumers will be purchasing a product that does not meet their expectations.
The new pooled method displays this tendency to a much lesser extent, thereby having a higher consumer accuracy value which approaches the one from the CV selected and separate least squares methods. 

Next, we turn to muscle$\times$cook combination B9 ($n=49$) which is a representative example of muscle$\times$cook combinations with few observations.
The proposed CV selected method has a clear advantage over both the classic pooled and separate least squares method.
The classic pooled method fails to discriminate in its star rating since it only returns $3^*$ predictions, a tendency that we consistently observe across other muscle$\times$cook combinations. 
The classic pooled method is largely unable to predict $4^*$ and $5^*$ ratings. For some muscle$\times$cook combination it might still achieve a relatively high accuracy but this occurs rather artificially due to the dominance of $2^*$ and $3^*$ observations in the data sets.
Separate least squares results in a greater variation in star predictions but both in the direction of under-prediction as well as over-prediction, with the latter representing a poor consumer outcome. 
The CV selected and new pooled method, in contrast, do generate higher accuracies by limiting their over-predictions though they still struggle to correctly predict the $4^*$ ratings as they only rarely occur.

Finally, we consider muscle$\times$cook combination A7 ($n=162$), which forms an interesting example of those rare classes that consist of a rather large proportion of  better than everyday ($4^*$) and premium quality $5^*$ samples.
All predictive methods perform similarly in terms of the accuracy measures.
Nonetheless, they do struggle to correctly predict these $5^*$ ratings, with this behaviour being most apparent for the classic pooled method.
Each method does classify the majority of such samples as $4^*$, thereby still leaving consumers with a purchased product that will exceed their expectations, as recognized in the high value of the consumer accuracy metric.
Nonetheless, such incorrect classifications do represent an opportunity cost for the processor.
While we make use of standard industry star thresholds, 
future work could consider tailoring the star thresholds to find optimal decision boundaries, but this is beyond the scope of this paper.

\section{Conclusion} \label{sec_conclude}
The lure of extracting maximum value from the carcase while keeping processing costs low has seen increased interest in boning room automation with a greater ability to differentiate product at a muscle level to pack for differentiated product lines. 
Increasingly sophisticated boning rooms along with new objective measurement technologies, such as X-ray imaging, CT scanning and hyperspectral cameras (e.g., \citealp{allen2021recent}), will drive the need for further method developments in the space of precision eating quality prediction. 
While more traits will be measured and more product differentiation is possible in boning rooms, to ensure maximal value extraction from each carcase, the measured traits must be reliably linked to consumer acceptance which underpins their willingness to pay.

This paper outlines a regularized predictive method that enables prediction of eating quality for individual muscles processed by a particular cooking method, thereby focusing on the customer satisfaction for an individual meal. 
We jointly model eating quality scores across various combinations of muscles and cooking methods, allowing for shared effects of important palatability predictors across muscle$\times$cook combinations. 
Importantly, the coefficients are regularized such that combinations where relatively few samples have been taken can ``borrow strength'' from similar combinations with more samples. 
This provides a balance between the two extremes of a single pooled model across all samples (low variance, high bias) and a separate model for each muscle and cook class (high variance, low bias). 

On a unique real data set from Meat Standard Australia, we have shown the utility of the proposed predictive method in achieving good prediction accuracy. 
Importantly, the method is data-driven and involves far less manual smoothing than the existing approach that relies heavily on prior knowledge and fitting many sub-models. 
Moreover, it delivers stable eating quality and star predictions across various muscle$\times$cook combinations that are either competitive or superior to the separate or classic pooled least squares method.
The accompanying coefficient paths are a key feature that enable the researcher to visualise the behaviour of coefficients for each muscle$\times$cook combination across the range of the tuning parameter, and thereby retain the ability to incorporate prior knowledge through the selection of the preferred model along the spectrum.

\begingroup
\bibliographystyle{asa}
\setstretch{0.05}
\linespread{0.5}
\bibliography{multiclassreg}
\endgroup
\end{document}